# A Fault Tolerant, Dynamic and Low Latency BDII Architecture for Grids

Asif Osman[1], Ashiq Anjum[2], Naheed Batool[1], Richard McClatchey[2]

[1]Pakistan Institute of Engineering and Applied Sciences (PIEAS), Islamabad, Pakistan

[2]Centre for Complex Cooperative Systems, University of the West of England, Bristol, UK

{Asif.Osman,, Ashiq.Anjum, Naheed.Batool, Richard.McClatchey}@cern.ch

## Abstract

The current BDII model relies on information gathering from agents that run on each core node of a Grid. This information is then published into a Grid wide information resource known as Top BDII. The Top level BDIIs are updated typically in cycles of a few minutes each. A new BDDI architecture is proposed and described in this paper based on the hypothesis that only a few attribute values change in each BDDI information cycle and consequently it may not be necessary to update each parameter in a cycle. It has been demonstrated that significant performance gains can be achieved by exchanging only the information about records that changed during a cycle. Our investigations have led us to implement a low latency and fault tolerant BDII system that involves only minimal data transfer and facilitates secure transactions in a Grid environment.

## 1. Introduction

The Berkeley Database Information Index (BDII) [1] plays a key role in any Grid infrastructure. The BDII database reflects the current status of the Grid resources that are available at a particular moment in time. Such information is normally collected and propagated by information services, which can be defined as "databases of resource attributes and metadata for system management and resource discovery" [2]. This information is used for tasks such as resource discovery, workflow orchestration, meta-scheduling, files transfer/cataloguing and security [3].

The current BDII model is hierarchical in nature. Agents running on each core node of a site collect the current status of the resources in value-attribute pairs [19], aggregate them and update the BDII on that particular site (the so called siteBDII). At a higher level in a Grid, the resources' data is fetched from all site BDIIs that are running at individual sites and is updated in a Grid wide BDII (the so-called *TopBDII*). This mechanism is depicted in figure 1. Thus having information about the resources and their status is critical for a fully functional Grid.

In the current model, the TopBDIIs in each cycle fetch resource information from their respective siteBDIIs. This information is fetched in an LDIF [13] [14] format and is stored in an ldap database [5]. This mechanism keeps the TopBDIIs updated with the latest information and as a consequence the TopBDIIs provide a global overview of the resources and services that are available at a given instance in a Grid. Normally the operation required to update or refresh entries in a database takes a few minutes for each cycle. An analysis of the LDIF data fetched during any two consecutive cycles from a siteBDII reveals that the values of only a few attributes undergo changes, while the remainder of the attributes remain the same (see table 1). Laurence Field et al., observe [7] that "97.8% of the changes are confined to 14 attributes only".

Another approach to studying the attributes' behaviour is to find the changes at record level in the LDIF datasets that are being fetched, rather than finding the attributes that have undergone changes in a cycle. This approach provides an effective mechanism for analysing the attribute changes and helps in proposing modifications in the BDII model for performance improvements. The approach also makes the implementation phase relatively simple, as will be discussed in the later sections.

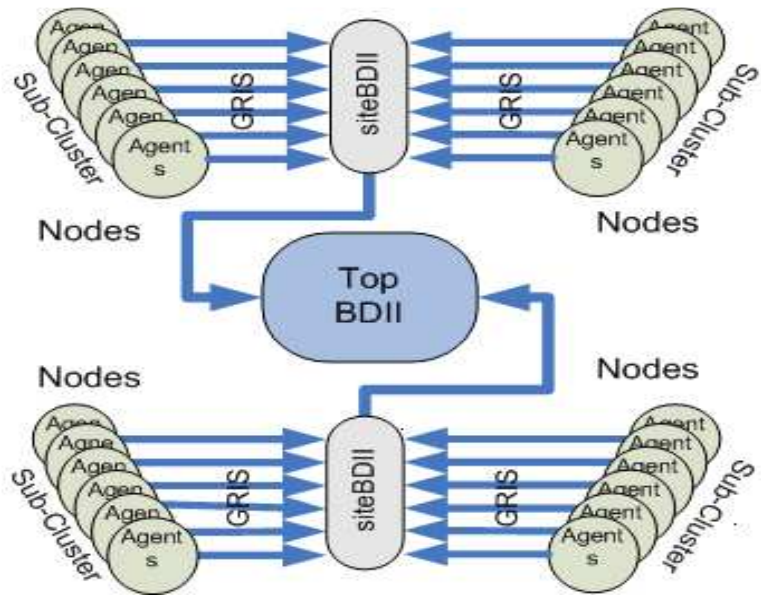

**Figure 1: Information gathering process in a TopBDII**

This paper discusses the findings of an analysis that has been performed on the fetched datasets and proposes modifications in the BDII model for performance improvement. Section 2 provides background to the proposed approach, whereas in section 3 the architecture and description of the proposed approach have been discussed. Section 4 describes the experimental framework that has been used to evaluate the proposed approach. Section 5 provides a detailed overview of the results that were achieved by implementing and executing the proposed algorithms. Section 6 gives an overview of the state of the art and section 7 concludes the findings and briefly presents the future directions in this work.

## 2. Background to the Proposed Approach

In the analysis described in this paper the same ldapsearch querying model (for collecting the LDIF datasets) has been used as is being employed in the current WLCG model [15]. The Worldwide LHC Computing Grid (WLCG) combines the computing resources of more than 323 computing centres in 55 countries, aiming to harness the power of 100,000 CPUs to process, analyze and store data produced from the LHC, making it equally available to all partners, regardless of their physical location. The trends of the attributes' changes and data redundancy were observed in the data that has been fetched from all sites, but being the most resource rich and representative site in the WLCG Grid, results have only been presented for the CERN-PROD production site [12]. The findings can also be applied to other Grid Information Services that employ models that are similar to BDII [18][19][20].

An experimental setup has been prepared to measure the differences in the datasets that have been fetched from a siteBDII during several consecutive cycles. The statistics generated from this experiment reveal that the number of records changed in each cycle, on average, is less than 2.37%. This led us to propose a more effective approach in which we only exchange the information in datasets that has undergone some change in a previous cycle. The details about the experimental setup and the results are described in the later sections of this paper. On the basis of these results, it is proposed to introduce a caching mechanism into the BDII model. The model will use the "differences" in the datasets instead of exchanging the whole datasets for updating the TopBDIIs. To dynamically utilize the caching mechanism in the current implementation, it is introduced as a plug-in since this will require very few modifications in the BDII implementation scripts. The following is a summary of the advantages that have been envisaged by implementing the proposed modifications:

- *Removing Data Redundancy*: The BDII schema [10] contains data for static and dynamic parameters. The static parameters undergo changes only at the time of resource reconfiguration in a Grid site. In the dynamic parameters, very few variations have been observed between datasets of any two consecutive cycles, due to the short time interval between the cycles. Developing a cache mechanism will eliminate the redundancies and improve the BDII performance.

- *Secure Data Transfer*: Currently the data transfer between a TopBDII and a siteBDII is triggered through ldapsearch, which is an insecure method of data transfer [3], since the data movement is in plain text. In the proposed model, the data from a local ldap database will be searched, at the same site where the siteBDII is collecting the data, and the "differences" will be calculated and stored locally in a file. This file can then be pulled by other remote TopBDIIs using secure protocols such as gridftp. This mechanism will help in providing secure data transfer without adding a new security layer in the BDII model.

- *Separating Retrieval and Remote Transfer Mechanisms*: The proposed model suggests that the data should be retrieved and stored in a file locally. This file should then be pulled by other remote TopBDIIs through secure protocols, as discussed in the previous paragraph. This mechanism will separate the search and the remote transfer operations, thus minimizing the probability of failures when compared to the data querying process using the ldapsearch command. In addition the new protocol for fetching the "differences" file may have built-in multi-streams/data recovery strategies. This feature will help in providing a swift transfer of the data and, in case of a failure, will provide some level of data recovery. Separating the retrieval and the transfer mechanisms has also another advantage in that a transfer protocol can be replaced with an improved one, as and when new releases or implementations of such protocols become available.

- *Checking Data Transfer Authenticity*: The data can get corrupted during transfer due to a number of reasons. In the current model there is no mechanism to verify the authenticity of the data that is being transferred. The proposed model will require the metadata to be encapsulated alongside the dataset. This metadata will complement the data, with some additional information, for improving the quality and authenticity of the data.

- *Data Compression*: Extra mileage can be drawn from the proposed scheme by implementing various compression methods on the files that are being exchanged. This mechanism will further reduce the data transfer time.

As a database model, BDII caters for both the storage and the retrieval of information. The approach in this paper will remain limited to proposing and implementing the improvements at the highest tier i.e. the TopBDII. At this level, significant amounts of data are exchanged, the redundancies in the data are quite high and the frequent data transfer processes consume much of the internet bandwidth in a Grid infrastructure. The proposed model has been introduced as a plug-in to the existing system. This scheme has been devised so that during the implementation phase, the proposed model should not disturb the current working of the model in the WLCG. The new scheme will add some overhead in finding the differences and then generating the ldif files, however, as discussed later in the results section, the advantages of the new scheme far outweigh the overheads introduced.

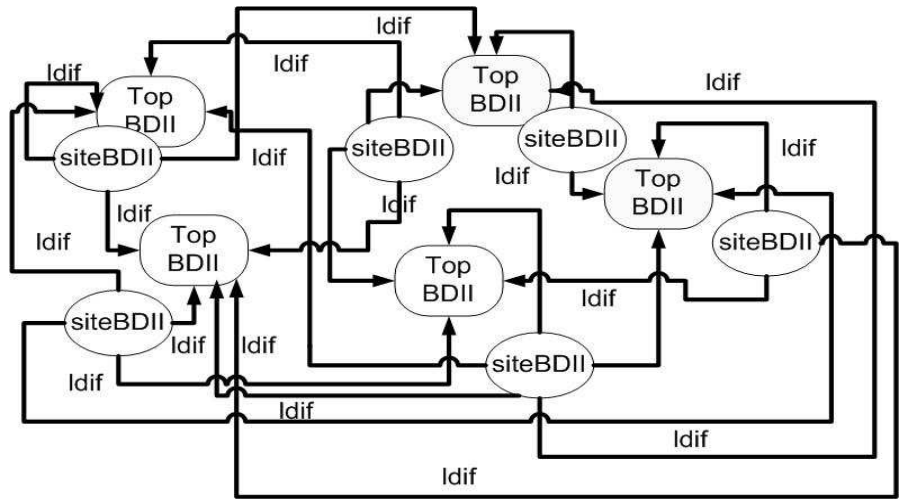

**Figure 2: ldif dataset exchanges between TopBDIIs**

## 3. Description of the Proposed Architecture

Before the proposed model is presented, it is useful to briefly discuss the current BDII approach. In the existing model, each TopBDII fetches datasets from its siteBDII servers that are available in the Grid infrastructure. A BDII script (the so called *BDDI-update*) is invoked during each cycle which sends queries to remote BDIIs, by making use of the URLs that have been retrieved from a configuration file. The outputs are stored in files that are uniquely named according to the site's abbreviated naming conventions. The naming conventions are also retrieved from the configuration file. The output files thus created are given the ldif extension. In order to achieve efficiency, the BDII script forks as many child processes as there are URLs in the configuration file. At the end of each cycle, the BDII script rebuilds the ldap database and its indices by merging all the data in the Glue Schema [10] format, using the ldif files that have been recently fetched [7]. It means that if N sites are running TopBDIIs in a Grid infrastructure and there are M siteBDIIs connected to the TopBDIIs, then M*(N-1) sets of data will be exchanged during each cycle. This mechanism is depicted in figure 2.

### 3.1. The proposed BDII model

During the data exchange and synchronization process between the TopBDIIs, as discussed earlier in the introduction, significant data redundancy has been observed. It is proposed that instead of exchanging the data as a full dataset, if only the changes in the data that occurred in a site during each cycle are exchanged, the performance of a Grid infrastructure will be improved. The proposed mechanism works in the manner now described.

To enable this "changes only" approach to function properly, it is necessary to locally calculate the differences between the datasets that have been fetched from each siteBDII. For this to happen, an algorithm needs to be implemented at every siteBDII, which queries the local resources in the same manner as it queries the remote siteBDII in the current model. The retrieved data should be stored in a file, hereafter called the ldif file. To establish the differences, this data should be compared with the dataset that was retrieved and saved in a previous cycle. The changes found are stored in another file, hereafter called the dif file. All TopBDIIs, which so far pull the full datasets, may now fetch these dif files. An up-to-date dataset can then be generated by patching the changes stored in the dif file to its corresponding dataset, that has been saved in an ldif file during a previous cycle. The generated datasets can then be used for updating their local databases in a similar way since they are being used in the current implementation. This means that the existing model will in no way be disturbed and a caching mechanism can be introduced for exchanging the ldif data. By getting rid of the major redundancies in the ldif data, it is expected that the sizes of the files, that contain only the differences in the datasets, will be small when compared to the full datasets, thereby significantly reducing the time required for their transfer. The proposed model is depicted in figure 3.

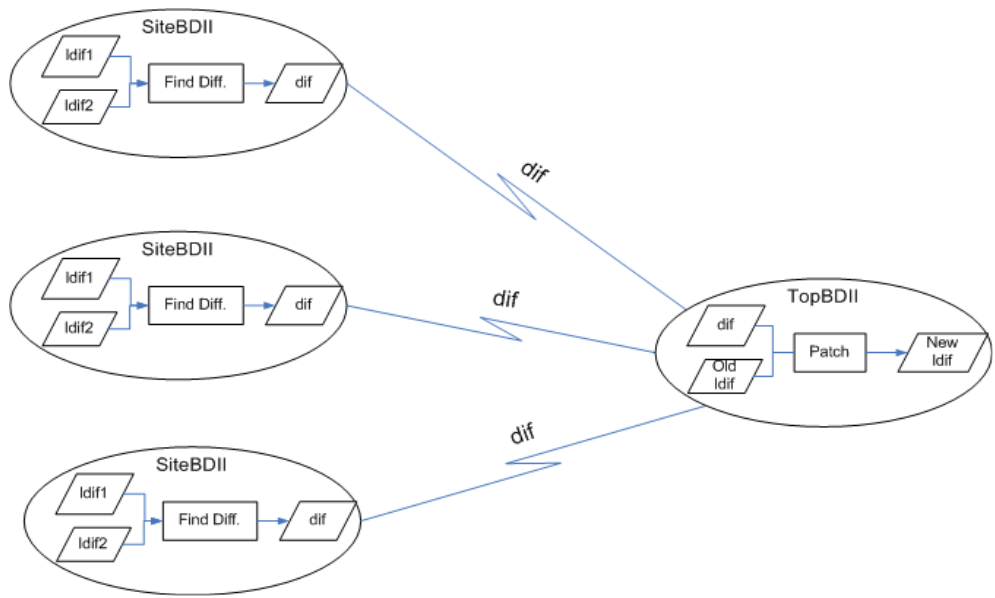

**Figure 3: Only changes in the parameters are exchanged in the proposed mode**

Some metadata needs to be wrapped up along with the differences dataset so that the goals defined in the introduction section of this paper can be achieved. This metadata will help in verifying the integrity and authenticity of the data and will contain the information that is required during the implementation phase. The metadata will be saved as an XML file, so that its different fields could easily be parsed.

A template for the wrapper file is given here:

```
<wrapperfile>
        <action> Value </action>
        <cycleno> Value </cycleno>
        <freshtime> Value </freshtime>
        <checksum> Value </checksum>
        <payload>
                records
        </payload>
</wrapperfile>
```

### 3.2. The proposed approach

As has been discussed in the previous sections, the "differences" are calculated at the same siteBDII where the data has been produced (acting as a server) and are fetched by all other TopBDIIs for their consumption (acting as clients). This means that the server's and client's parts will be working asynchronously and a server side script will start producing dif files as soon as a siteBDII is in operational mode. However, the timing for fetching the files will differ from one site to another, depending on when the execution starts at a particular site. To ensure synchronisation, it is proposed that a server side script must always produce two types of files during each cycle; one file storing a full dataset and the other file storing the differences between the datasets. A cycle number is assigned to each file that has been produced in the cycle. The same cycle-number needs to be stored under the attribute <cycleno> for both ldif and dif types of the wrapper files. The checksum of the full dataset file is also calculated and is stored in the wrapper file. Refer to the XML template above for all the attributes discussed here.

When a script running at a TopBDII as a client starts retrieving data from a remote siteBDII for the very first time, it begins by fetching the file containing the full dataset and saves it in a ldif file. This file will serve as the base LDIF dataset. After waiting for the <freshTime>, it fetches a dif file with a string "cycle-no+1" that is appended in the file name. The data thus found will be used for the generation of the current cycle's LDIF data by patching it to the saved ldif file that was received during the previous cycle. This process can be repeated, by fetching dif files, doing the patching and obtaining the data for the current and next cycles. The checksum of the generated file should be compared with the checksum value contained in the wrapper to ensure the validity of the generated data. An <action> value "Initialize" has been used to instruct the client that it should restart itself. It is required in the case where a server side script fails due to the reasons discussed in the following paragraphs. In normal cases, this attribute contains a "Normal" value.

If at any stage during the operation it is found that the cycle-number of a server side differs from that of a client side then the two sides have become out of sync. In this case, it should follow one of the solutions discussed under the failure cases. By default, the daemon running on the server side is invoked as a service with an action-value of "Normal". In the case of a problem, a cron job will investigate and invoke it with an action-value of "Initialize" which is an indicator to all clients that the server side script has been restarted. The discussion on using a cron job as a test script is detailed in the following paragraphs.

The configuration file will contain the sites's abbreviated names (abbrev-names), delta-time and query details. The abbrev-names should be the same as those found in the configuration file on the client side. The delta-time is the wait time between two queries. A query is formed according to a local site_bdii configuration, for example, for CERN-PROD site; it will look like the following:

*ldapsearch -x -LLL -h prod-bdii.cern.ch -p 2170 -b mds-vo-name=CERN-PROD,,o=grid '(|(objectClass=GlueSchemaVersion)(objectClass=GlueTop))'*

To make the system robust, a script has been implemented that keeps track of the status of a process. It has been proposed to run this script as a cron job. This type of script should be running both on the server and on the client side. On its invocation, it will check whether the client and server side scripts are behaving normally. The criterion for checking the status is to find that the file to be created during each cycle has been created in time. This is done by comparing its date stamp with the system date. If the difference is beyond some tolerance limit, a cron job will investigate and then re-execute the failed daemon.

The proposed scheme may have shortcomings. Some of the possible issues and their remedies are the following.

   a) a server side script may fail
   b) a client side script may fail
   c) a checksum may reveal that the data has been corrupted during the transfer/generation
   d) a retrieval side may fail to retrieve dif files during one or more cycles

The solution for the first of these problems is to restart the server side script, by putting the string "Initialize" in the <action> section of the wrapper file. This notifies the clients that the server side script has been restarted and that they need to readjust their values. All the client sides on sensing the indicator should retrieve an ldif file, instead of a dif file, and should restart themselves as they did for the first cycle. To address the other issues in the list, it is again recommended to restart the client side script; it should start operating in the same way as it did after receiving the "Initialize" indicator. For restarting the client or server side scripts, another script has been written, which is invoked by cron. This assesses the status of currently running scripts and re-runs a new copy with initial conditions, after killing the old processes, just in case a script is malfunctioning. Another problem that has been observed is related to the status of a siteBDII. If a siteBDII is down, the script for updating a ToBDII keeps on waiting until a time-out occurs. It is thus recommended to ping all the sites occasionally and skip those which are inaccessible for any reason.

## 4. Experimental Setup

Due to security issues and access privileges, it may not be possible to deploy the proposed system on all sites in the WLCG Grid. It may also not be desirable to deploy a prototype of the proposed approach in a production Grid infrastructure since it could disturb its operation. Therefore, the sample data for this research study was retrieved from the CERN-PROD site and was then transferred to a local setup for further studies. The data was captured for 100 cycles and each cycle had an interval of 5 minutes. For the experimental setup, two scripts have been implemented to calculate the statistics mentioned in this paper. The differences between the data of consecutive cycles were calculated and are plotted as shown in figure 4.

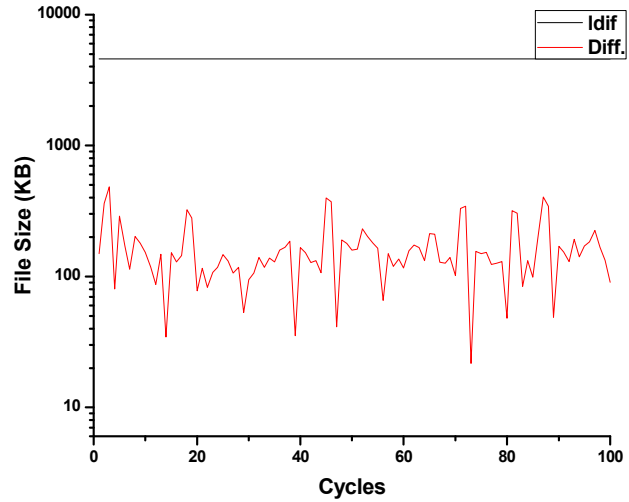

**Figure 4: Difference between data of consecutive cycles**

The experiment was repeated at different times and it was found that the number of records that were changed in a cycle were less than 2.37% of the total ldif records. The files that contain changes in the datasets are quite small in size when compared to the ldif files that contain the full datasets. Zipping the dif files yield even lighter file sizes (see figure 5). It is obvious that exchanging the comparatively smaller files will reduce the amount of data transfer across wide area networks in Grids, save bandwidth and reduce the failure rates, thus improving the performance of BDII [8][17].

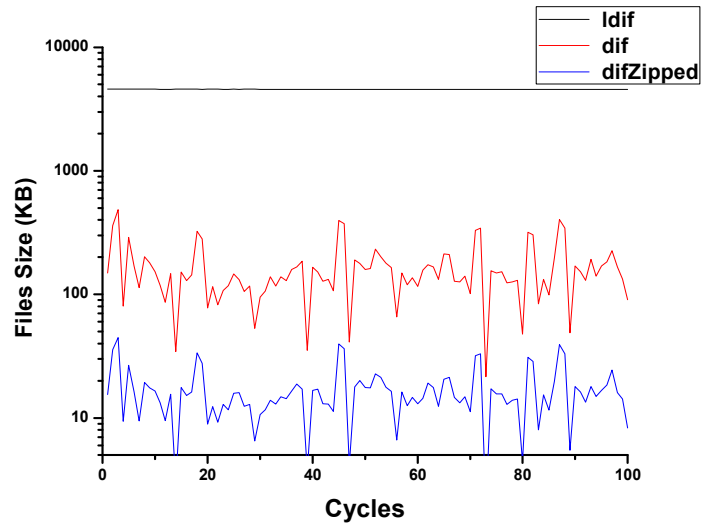

**Figure 5: ldif, difference and zipped difference file sizes**

The test scripts that have been written for this study do not contain any logic for generating wrapper and meta data parts, which are actually required at the implementation time. The wrapper feature can easily be incorporated in the scripts at the deployment time. The dif files have been processed further to calculate the different types of statistics presented in this paper and are discussed further in the results and discussion section. To check the validity of the generated files, they were compared with an original copy of the datasets by using the procedure now explained.

We started with an ldif file that contained a full dataset which was taken as the 0th cycle data. It was patched with the "differences" data of the following cycle, thus generating a full dataset for the next cycle. The generated file was compared with its corresponding original file and was found compatible at the bit level. The process was repeated for the whole sample data and complete agreement was verified.

In the proposed approach, it has been considered that the model should be implemented as a plug-in to the existing BDII model. In the existing model, a script (the so-called *bdii-update*) is executed and its configuration parameters are kept in a file called bdii.conf [16]. The script, using ldapsearch, keeps on fetching the data from the siteBDIIs in a Grid and stores them in files. At the end of a cycle, these files are merged and used for updating the TopBDII. Under the new scheme, a plug-in has been introduced into bdii-update by making a few minor modifications in the existing code. The plug-in implements a caching mechanism to carry out the required processing at local sites. After the modifications the system does not fetch the data directly from the remote siteBDIIs. The proposed system can now consume the cache data and resumes its processing in a normal way as before. Using this approach the proposed scheme can be introduced seamlessly in the production BDII's. To control the enabling/disabling of the plug-in, a new attribute "PLUGIN" has been introduced in the bdii-update.conf file. In a routine operation, bdii-update will be running as a normal legacy code, with PLUGIN set to OFF. By setting the value of the PLUGIN attribute to ON, it starts working according to the new scheme. This approach will help in creating a dynamic plug-in for the new scheme, without disturbing the current working of a Grid. Under the new framework, it is mandatory to keep on executing the server side software so that it should always be able to provide the dif files to the sites that have adopted the proposed model.

## 5. Results and Discussion

The worldwide LHC Computing Grid (WLCG) [11][15] project was initiated with the aim of developing the Grid for the Large Hadron Collider (LHC) at CERN, Geneva, Switzerland. The LHC is expected to produce around 15 Peta bytes of data annually. This data is replicated around the world to enable thousands of physicists to carry out their analyses from their home institutes. The WLCG is assigned the task of selecting and deploying appropriate middleware – the software that provides seamless access to the computing Grid, and controls and monitors its performance. It is then ready to provide an environment for job execution to hundreds of global users [4]. Currently 323 grid sites located in 55 countries are contributing computing resources to this project. At the time of this paper, the WLCG grid is functional with computing nodes of more than 73,000 CPU's and a storage capacity of more than 195 PB. Some of these sites are interconnected via multiple 10GB WAN links. Through the Grid these resources are available to its users, as if it is virtually a single resource.

As stated earlier in this paper, data was fetched from all siteBDIIs of the WLCG production Grid and was stored in files in the LDIF format. At the time of experiment, it was found that almost 323-siteBDII servers and a fewer number of TopBDIIs are deployed in WLCG [20]. There are almost as many TopBDIIs as the number of sites (with a few exceptions as a few sites are using TopBDIIs of other sites). Figure 6 shows the graph of these file sizes. The site with the highest peak belongs to CERN-PROD. The ldif files were fetched from all siteBDIIs for one cycle and their total size sums to 66MB for the sample data. Using the familiar M*(N-1) formula explained in section 2, we can calculate that 66M*322 = 21GB data is transferred in just one cycle.

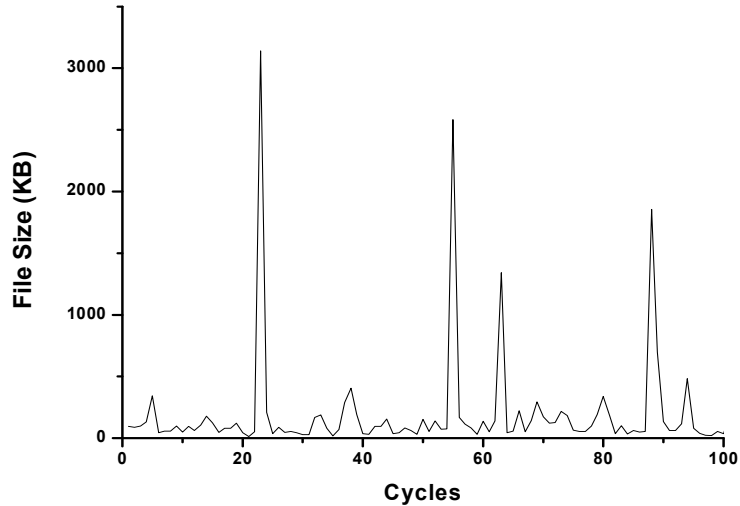

**Figure 6: LDIF files from all sites**

There are almost as many TopBDIIs as the number of sites (with a few exceptions as a few sites are using TopBDIIs of other sites). Figure 6 shows the graph of these file sizes. The site with the highest peak belongs to CERN-PROD. The ldif files were fetched from all siteBDIIs for one cycle and their total size comes out to be 66MB for the sample data. Using the familiar M*(N-1) formula explained in section 2, we can calculate that 66M*322=21GB data is transferred in just one cycle.

A frequency distribution table that has been prepared from the ldif data is shown in table 1. The data clearly shows that 99.78% of the most frequently changing parameters are confined to only a few attributes. Almost similar statistics were calculated by Field et al. in their paper [7] and are reproduced later in table 4 for comparison.

**Table 1: Most Frequently Changing Attributes**

| Attributes | Freq. (%) | Attributes | Freq. (%) |
|---|---|---|---|
| GlueCEStateEstimatedResponseTime | 27.49 | GlueCEPolicyAssignedJobSlots | 0.25 |
| GlueCEStateWorstResponseTime | 26.17 | GlueCEInfoTotalCPUs | 0.25 |
| GlueCEStateFreeJobSlots | 11.51 | GlueSEUsedOnlineSize | 0.16 |
| GlueCEStateFreeCPUs | 10.08 | GlueSESizeFree | 0.16 |
| GlueCEStateTotalJobs | 8.33 | GlueServiceDataKey | 0.08 |
| GlueCEStateRunningJobs | 7.65 | GlueSchemaVersionMinor | 0.05 |
| GlueCEStateWaitingJobs | 3.01 | GlueSchemaVersionMajor | 0.05 |
| GlueServiceStartTime | 1.09 | GlueTop | 0.05 |
| GlueSAStateAvailableSpace | 0.83 | GlueSchemaVersion | 0.05 |
| GlueServiceStatusInfo | 0.75 | GlueKey | 0.05 |
| GlueSAUsedOnlineSize | 0.54 | GlueChunkKey | 0.04 |
| GlueSAFreeOnlineSize | 0.54 | GlueServiceDataValue | 0.04 |
| GlueSAStateUsedSpace | 0.27 | GlueServiceData | 0.04 |
| GlueCEPolicyMaxRunningJobs | 0.25 | **Total:** | **99.78** |

A further investigation into the "differences" files reveals that only a small fraction of the total number of records in the ldif datasets underwent changes. Even the records that contain changes have several duplicate entries. As a consequence the unique records have been identified and the statistics are plotted in figure 7. From this information, we can conclude that a significant redundancy still remains in the files and further gains can be achieved if these dif files are zipped before being exchanged.

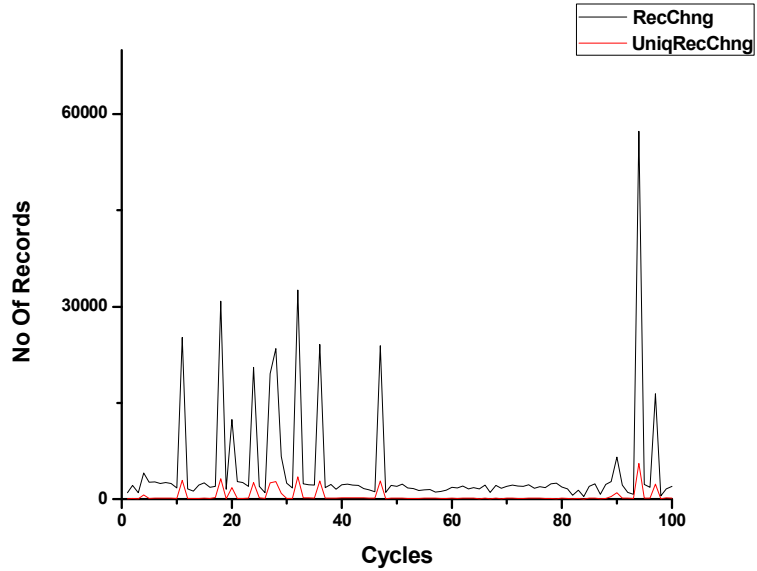

**Figure 7: Number of changed records/Unique records**

The records that underwent changes cannot be used in the statistical analysis, because they are not in a format that can be used for patching. However, the differences obtained using the diff commands in Linux are in a format that can be further utilised. This format carries an overhead, which helps in generating the n+1th cycle ldif file, by applying the "differences" on the nth cycle ldif file. The overheads include a slight increase in the sizes of the dif files, the additional time taken during the execution of the scripts for finding the differences at the server side and the additional time spent in the unzipping and patching processes to regenerate the ldif files at the client site. The unzipping and patching process in particular consumes more time as this has to be done for all the dif files. As an illustration, the times spent on gzipping, and then serially generating the files by the unzipping and patching operations on 100 files are shown in table 2:

**Table 2: The time spent on gzipping, unzipping and patching**

|  | Gzipping | Unzipping and Patching |
|---|---|---|
| Real Time | 0m0.655s | 0m22.802s |
| User Time | 0m0.451s | 0m2.991s |
| System Time | 0m0.041s | 0m3.527s |

These times have been calculated on a 32-bit machine with a 3.2GHz processor. The time taken can be reduced by performing these tasks in parallel, using a multithreading approach.

The sizes of the dif files in zipped as well as unzipped forms are plotted in figure 8. The data transfer activity between TopBDIIs is accomplished through ldapsearch, which is currently insecure. In the proposed system, "globus-url-copy" has been used for the data transfer, which will increase the efficiency and security of the transfer protocol. If a protocol that is being used has an inherent capability of multi-streaming, it will further improve the data transfer efficiency. The peaks in figures 7 and 8 look different because they represent different data sets. In figure 7 the numbers of changed records have been plotted whereas figure 8 represents the dif file sizes in zipped and unzipped formats.

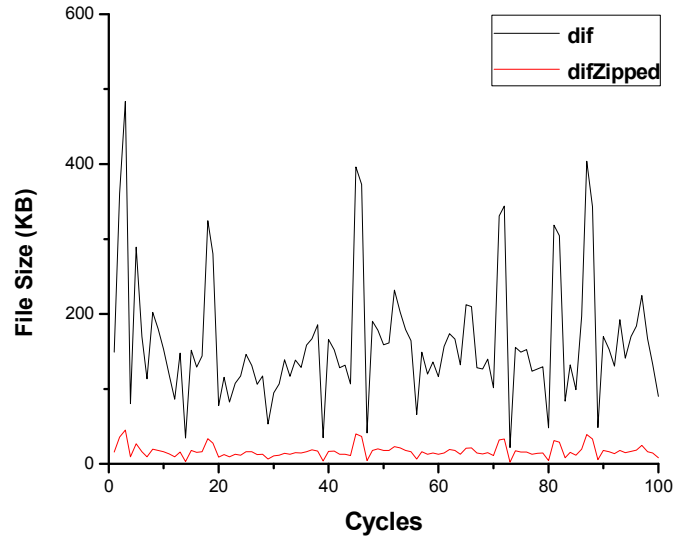

**Figure 8: Changes between the consecutive cycles (normal/zipped)**

For comparison, the sizes of the ldif files that contain the original data are shown in figure 9.

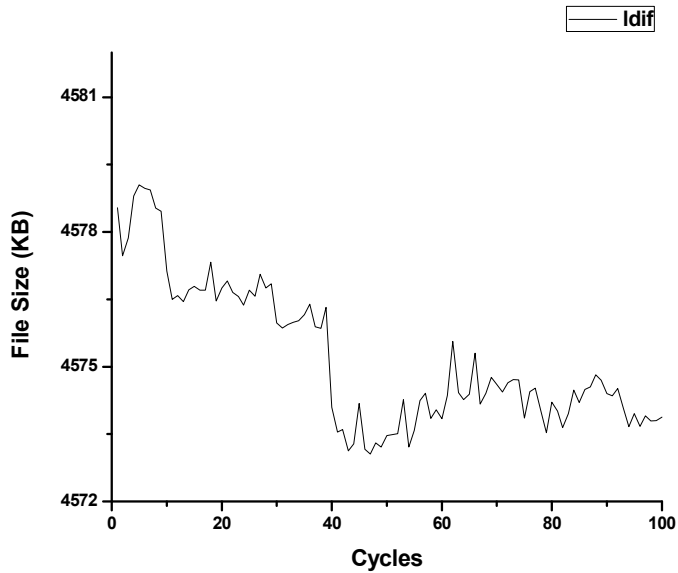

**Figure 9: File sizes of raw ldif datasets retrieved from the CERN_PROD site**

The trends in figure 9 show that the dataset size retrieved from the CMS-PROD site during each cycle remains almost the same (~4.57MB). These graphs suggest that the response of the BDII system, as a result of the proposed changes, is such that it is a good candidate for the addition of a cache mechanism in its architecture. As discussed in the proposed model section, the changes as well as the metadata are saved in the dif files. The sizes of these files for a run of 100-cycles are shown in figure 8. Comparing figures 8 and 9 shows that the implementation of the proposed model has reduced the dataset size to almost 27KB per cycle from 4.57MB, which is a significant optimization. If this saving is considered in the light of the refresh-time and the available sites in the Grid, the optimization achieved could be around 99.40%, which is difficult to ignore for optimal Grid operations.

We are also expecting similar reductions in the data transfer from other sites because all of the Grid sites operate under similar conditions, as far as the BDII model is concerned. The proposed scheme becomes more significant in the light of the fact that the trends in almost all Grid infrastructures show that the number of sites and the number of jobs executing are ever increasing [7]. This phenomenon contributes directly towards the amount of the ldif data that is being exchanged in a Grid. During the operation of the Grid site PAKGRID-LCG2, it was observed that the ldif data retrieved in a cycle from the CERN_PROD site increased from ~1.5MB to ~4.5MB during the period 2007-2010, as a greater number of sites joined the WLCG during this period.

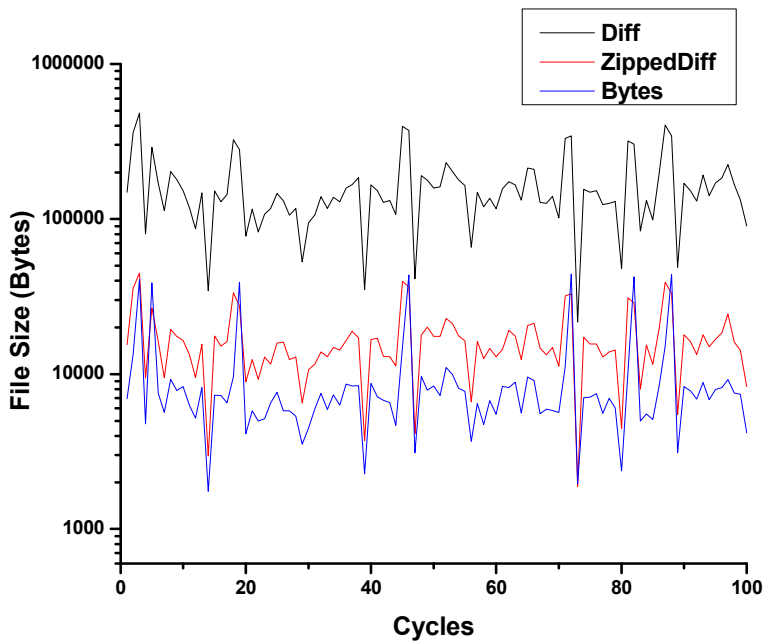

**Figure 10: Bytes level changes comparison with zipped dif file sizes**

To provide comprehensive statistics about the total number of records found in the raw ldif files, the number of records that have been changed in each cycle, and the bytes level changes are shown in figure 10. To identify the byte level changes, the values of the attributes containing the changed records were extracted and plotted by counting the number of bytes involved in the change. These numbers can be taken as a theoretical limit, which suggests how much redundancy can be removed. It is obvious from the nature of the data that the dif files contain many duplicate records. The attributes are taken from the same schema and are repeatedly populated, so there remains a lot of redundancy in the data.

The results demonstrate that the compression operation will further reduce the data transfer size. This fact is also verified in the results that have been presented in the figures in this section. The reduction in the data that is being transferred will obviously improve the performance of BDII. The data transfer operation will consume less bandwidth, which means that even the sites that do not have good internet connectivity can maintain their own TopBDIIs. Maintaining a local TopBDII at each site will reduce the retrieval time and the probability of failures. It will also help in load balancing. Normally a BDII database performs many more retrieval operations than updates (due to an ldap server), so it is expected that having more TopBDIIs in a Grid will have a profound effect on the performance of a Grid. The efficiency of a TopBDII depends on how much up-to-date data it contains at a given instant and this is very much dependent upon the refresh time. After the implementation of the proposed scheme, it is now possible to reduce the refresh time interval. Due to the smaller sizes of the files that need to be transferred, this will have a very profound effect on the resource utilization where TopBDII's are hosted. To save time and to speed up the operations, the TopBDII script spawns many child processes that fetch the information from all siteBDIIs in parallel. Due to the smaller file sizes, these child processes need small buffers and also collapse in relatively shorter periods of time, thus releasing the resources promptly. This has also helped in improving the scalability of the BDII model.

During the data fetching operation from the CERN_PROD site, it was observed that the ldapsearch client failed with the error "Can't contact LDAP server (-1)" for almost 14-16% of the cycles, providing partially retrieved datasets. The percentage of failures may be even higher at peak loads on a site. It was also observed that the LDAP server in a BDII does not provide any recovery mechanism in case of a failure. The same data when collected in CERN (locally) reduced the failure rate to almost zero. It is expected that a protocol used for fetching the files should have a built-in mechanism for a fail-safe data transfer. The proposal made in this study will provide the remedy for testing the failures and improving the chances of recovery. Moreover, the checksum value will verify that the data at the producer and consumer sides are in agreement. If some differences are found, the original ldif file can be downloaded and the same mechanism can be carried out again. The patching operation may complain if a dif file does not contain the proper data and thus should store the rejected records in another file. Some additional features can also be implemented in the code that runs at the client side for checking the integrity of the data. For example, it can retry if globus-url-copy fails to download the data. The same script keeps on checking the update operation on the regenerated data. If these files are not updated within a tolerable time limit, they can be downloaded again.

The achieved optimization comes at a cost. The individual BDII servers have to do additional work to identify the changes and some time is always spent on data encoding and decoding operations. Once a dif file has been fetched, a site has to do more work in reconstructing the ldif files from the decoded data. This means that the operations will add more complexity at the TopBDII and siteBDII levels in Grid sites. However, the savings achieved in the internet bandwidth, the data transfer time, in achieving safe and secure data transfers and in reducing the probability of service failures far outweigh the overheads. The overheads that are involved in providing additional activities such as coding/decoding were calculated and, as shown in table 3, they are found minimal when compared to the gains.

The data presented in table 3 suggests that a file of 27K bytes requires a fraction of the time when compared to downloading a file of 4.57M bytes. Moreover, the overheads incurred for processing the data for the cache mechanism seem quite tolerable. Taking into account the case that all TopBDIIs are fetching the same amount of data, the savings become quite significant. These statistics were calculated on a Linux box with a Intel Pentium(R) 4 and having a CPU of 3.40GHz.

**Table 3: Statistics obtained during the validation of the proposed model**

| Actions | | Remarks |
|---|---|---|
| Size of raw data retrieved | 4.5MB | |
| Size after coding/zipping | 27KB | Reduction > 99% |
| Time taken during coding and zipping | <0.1s | |
| Time taken during unzipping and generation | <0.1s | |
| Total overhead (time-wise) | <0.2s | |
| Diff. file transfer time | <2s | |

# 6. Related work

Field et al. in their paper [7] calculated the mutability in the BDII data. The conclusions drawn in the paper show a direct correlation with the work presented in this study. The frequency distribution for the frequently changing attributes has been reproduced in table 4. It is worth noting that the set of attributes and their frequencies measured in this study and the one calculated by Field et al. do differ. In our opinion, these values will always differ and depend upon the instance when the data is collected and the scope of the data that is being analyzed. Our dataset is confined to the CERN-PROD production site whereas Field's et al. have collected it from the whole WLCG for a week, approximately [7].

**Table 4: Mutability in BDII data**

| Attribute | Percentage |
|---|---|
| GlueCEStateTotalJobs | 9.41% |
| GlueCEStateFreeCpus | 9.52% |
| GlueSAStateUsedSpace | 5.38% |
| GlueCEStateFreeJobslots | 19.36% |
| GlueCEStateWorstResponseTime | 11.79% |
| GlueSASateAvailableSpace | 6.57% |
| GlueCEStateEstimatedResponseTime | 12.50% |
| GlueCEStateRunningJobs | 7.90% |
| GlueCEInfoTotalCpus | 4.67% |
| GlueCEStateWaitingJobs | 6.37% |
| GlueCEPolicyAssignedJobSlots | 0.90% |
| GlueServiceStartTime | 0.71% |
| GlueSAUsedOnlineSize | 1.34% |
| GlueSAFreeOnlineSize | 1.37% |

The proposal made in this study is consistent with the approach adopted in [7], the only difference being that the referenced paper calculates mutability and suggests its use at the tier level 1 during implementation of the existing model, whereas our approach proposes a caching mechanism that needs to be introduced as a plug-in at the tier level 2. As discussed earlier, in such a dynamically changing system, one cannot predict which set of attributes will always be involved as an agent of change. The set of attributes being monitored can vary from time to time. Therefore any solution provided must be sufficiently generic to cater for the dynamically changing situation in the system. The modifications proposed in this paper, for finding the changes at records level, appear to meet the requirements of this criterion.

Astalos et. al. in their paper [6] state that each TopBDII should ask for an entire site info, thus generating a huge network load. A BDII could be limited to changed entries only. A BDII database could be reloaded after each rebuild, indices would need to updated and the hierarchy could be improved at the top level. But they suggest a solution which requires an overhaul of the entire BDII system for its implementation. In our opinion, it may not be feasible to introduce a completely new information system in an existing production system since such changes will affect many other components in the Grid.

## 7. Conclusions and Future work

This paper presented a performance aware model of BDII. It has been demonstrated that significant performance gains can be achieved in a Grid infrastructure with the adoption of the proposed model. The implementation includes mechanisms for the information retrieval from the BDII databases, the algorithm for finding the changes for a particular number of cycles and decoding it in a format suitable for the caching. The algorithm implementation can also generate ldif files and update the BDII database using the decoded data. These generated ldif files were found to be 100% bit compatible with the original ldif datasets, when compared with the data of 100-cycles having an interval of 5 minutes between them. The algorithm used in the proposed model dictates that the data generated during the 100th iteration depends on the 99th iteration, which itself depends on the 98th iteration and so on. The compatibility of the data generated in the 100th iteration with the original data, which was retrieved directly by ldapsearch after the 100th cycle, proves the reliability of the proposed algorithm.

This exercise is sufficient as a proof-of-concept that the proposed solution will work for all sites in a Grid, however, the plug-in does need to be fully integrated and rigorously tested for a production environment such as the WLCG. For the collection of the statistics and the generation of the ldif files, this test case was executed for one site (CERN) only. For real tests, it needs to be launched on all TopBDIIs participating in a Grid. This is not currently possible due to limited accessibility on account of security issues and operational limitations. However, the provision of the plug-in on/off feature enables the sites to adopt the proposed BDII system at their convenience. It will help them in a smooth transition when they want to adopt the new model, without disturbing the working of a Grid. All the components that have been introduced in the proposed model are available as an open source project, can be selected on merit and performance and can be replaced or upgraded as a new version or an equivalent component becomes available.